\documentclass[11pt]{article}
\begin{document}
\title{{\bf{\Large Hawking black body spectrum from tunneling mechanism}}}
\author{
 {\bf {\normalsize Rabin Banerjee}$
$\thanks{E-mail: rabin@bose.res.in}},\, 
 {\bf {\normalsize Bibhas Ranjan Majhi}$
$\thanks{E-mail: bibhas@bose.res.in}}\\
 {\normalsize S.~N.~Bose National Centre for Basic Sciences,}
\\{\normalsize JD Block, Sector III, Salt Lake, Kolkata-700098, India}
\\[0.3cm]
}

\maketitle

\begin{abstract}
   We obtain, using a reformulation of the tunneling mechanism, the Hawking black body spectrum with the appropriate temperature for a black hole. This is a new result in the tunneling formalism of discussing Hawking effect. Our results are given for a spherically symmetric geometry that is asymptotically flat.
\end{abstract}

{\textbf{\textit {Introduction}}}:
    After Hawking's observation \cite{Hawking} that black holes radiate, there were several approaches \cite{Gibbons,Christensen,Paddy,Wilczek,Robinson,Rabin} to study this effect. A particularly intuitive and widely used approach is the tunneling mechanism \cite{Paddy,Wilczek}. The essential idea is that a particle-antiparticle pair forms close to the event horizon which is similar to pair formation in an external electric field. The ingoing mode is trapped inside the horizon while the outgoing mode can quantum mechanically tunnel through the event horizon. It is observed at infinity as a Hawking flux. So this effect is totally a quantum phenomenon and the presence of an event horizon is essential. However, in the literature \cite{Paddy,Wilczek,Majhi1,Majhi2,Majhi3,Majhi4,Singleton}, the analysis is confined to obtention  of the Hawking temperature only by comparing the tunneling probability of an outgoing particle with the Boltzmann factor. There is no discussion of the spectrum. Hence it is not clear whether this temperature really corresponds to the temperature of a black body spectrum associated with black holes. One has to take recourse to other results to really justify the fact that the temperature found in the tunneling approach is indeed the Hawking black body temperature. In this sense the tunneling method, presented so far, is incomplete.

    In this paper we rectify this shortcoming. Using density matrix techniques we will directly find the spectrum from a reformulation of the tunneling mechanism. For both bosons and fermions we obtain a black body spectrum with a temperature that corresponds to the familiar semiclassical Hawking expression. Our results are valid for black holes with spherically symmetric geometry. Finally, we show the connection of our formulation with usual tunneling formulations \cite{Paddy,Wilczek} by exploiting the principle of detailed balance.

{\textbf{\textit {General formulation}}}:
   Consider a black hole characterised by a spherically symmetric, static space-time  and asymptotically flat metric of the form,
\begin{eqnarray}
ds^2=F(r)dt^2-\frac{dr^2}{F(r)}-r^2d\Omega^2
\label{1.01}
\end{eqnarray}
whose event horizon $r=r_H$ is defined by $F(r_H)=0$. For discussing Hawking effect by tunneling, the radial trajectory is relevant \cite{Paddy,Wilczek}. We therefore consider only the $(r-t)$ sector of the metric (\ref{1.01}).

   Now consider the massless Klein-Gordon equation $g^{\mu\nu}\nabla_\mu\nabla_\nu\phi=0$ which, in the ($r-t$) sector, reduces to, 
\begin{eqnarray}
-\frac{1}{F(r)}\partial^2_t\phi+F^{'}(r)\partial_r\phi+F(r)\partial^2_r\phi=0
\label{scalar}
\end{eqnarray}
in the black hole space-time (\ref{1.01}). Taking the standard WKB ansatz
\begin{eqnarray}
\phi(r,t)=e^{-\frac{i}{\hbar}S(r,t)}
\label{1.22}
\end{eqnarray}
and substituting the expansion for $S(r,t)$
\begin{eqnarray}
S(r,t)=S_0(r,t)+\sum_{i=1}^{\infty}\hbar^iS_i(r,t)
\label{expansion}
\end{eqnarray}
in (\ref{scalar}) we obtain, in the semiclassical limit (i.e. $\hbar\rightarrow 0$),
\begin{eqnarray}
\partial_tS_0(r,t)=\pm F(r)\partial_rS_0(r,t)
\label{1.23}
\end{eqnarray}
This is the usual semiclassical Hamilton-Jacobi equation \cite{Paddy,Majhi2} which can also be obtained in a similar way from Dirac \cite{Majhi3} or Maxwell equations \cite{Majhi4}. Also, this equation is a natural consequence if the chirality (holomorphic) condition on the scalar field with the WKB ansatz (\ref{1.22}) is imposed with the $+(-)$ solutions standing for the left (right) movers \cite{Majhi5}.

  Now since the metric (\ref{1.01}) is stationary, it has a timelike Killing vector. Therefore we choose an ansatz for $S_0(r,t)$ as
\begin{eqnarray}
S_0(r,t)=\omega t+{\tilde{S}}_0(r)
\label{1.24}
\end{eqnarray}
where $\omega$ is the conserved quantity corresponding to the timelike Killing vector. This is identified as the effective energy experienced by the particle at asymptotic infinity. Substituting this in (\ref{1.23}) a solution for ${\tilde{S}}_0(r)$ is obtained. Inserting this back in (\ref{1.24}) yields,
\begin{eqnarray}
S_0(r,t)=\omega (t\pm r_*);\,\,\,\, r_*=\int\frac{dr}{F(r)}
\label{1.25}
\end{eqnarray}

    For further discussions it is convenient to introduce the sets of null tortoise coordinates which are defined as,
\begin{eqnarray}
u=t-r_*,\,\,\, v=t+r_*.
\label{1.02}
\end{eqnarray}
It is important to note that expressing (\ref{1.25}) in these coordinates, defined inside and outside the event horizon, and then substituting in (\ref{1.22}) one can obtain the right and left modes for both sectors:
\begin{eqnarray}
&&\Big(\phi^{(R)}\Big)_{\textrm{in}}=e^{-\frac{i}{\hbar}\omega u_{\textrm{in}}};\,\,\, \Big(\phi^{(L)}\Big)_{\textrm{in}}=e^{-\frac{i}{\hbar}\omega v_{\textrm{in}}}
\nonumber
\\
&&\Big(\phi^{(R)}\Big)_{\textrm{out}}=e^{-\frac{i}{\hbar}\omega u_{\textrm{out}}};\,\,\, \Big(\phi^{(L)}\Big)_{\textrm{out}}=e^{-\frac{i}{\hbar}\omega v_{\textrm{out}}}
\label{1.251}
\end{eqnarray}

   Now in the tunneling formalism a virtual pair of particles is produced in the black hole. One member of this pair can quantum mechanically tunnel through the horizon. This particle is observed at infinity while the other goes towards the center of the black hole. While crossing the horizon the nature of the coordinates changes. This can be accounted by working with Kruskal coordinates which are viable on both sides of the horizon. The Kruskal time ($T$) and space ($X$) coordinates inside and outside the horizon are defined as \cite{Ray},
\begin{eqnarray}
&&T_{\textrm{in}}=e^{K(r_*)_{\textrm{in}}} ~{\textrm{cosh}}(Kt_{\textrm{in}});\,\,\,X_{\textrm{in}}=e^{K(r_*)_{\textrm{in}}} ~{\textrm{sinh}}(Kt_{\textrm{in}})
\nonumber
\\
&&T_{\textrm{out}}=e^{K(r_*)_{\textrm{out}}} ~{\textrm{sinh}}(Kt_{\textrm{out}});\,\,\,X_{\textrm{out}}=e^{K(r_*)_{\textrm{out}}} ~{\textrm{cosh}}(Kt_{\textrm{out}})
\label{Krus1}
\end{eqnarray}  
where, as usual, $K=\frac{F'(r_H)}{2}$ is the surface gravity of the black hole. These two sets of coordinates are connected by the relations,
\begin{eqnarray}
t_{\textrm{in}}\rightarrow t_{\textrm{out}}-i\frac{\pi}{2K};\,\,\,\, (r_*)_{\textrm{in}}\rightarrow(r_*)_{\textrm{out}}+i\frac{\pi}{2K}
\label{Krus2}
\end{eqnarray}
so that, with this mapping, $T_{\textrm{in}}\rightarrow T_{\textrm{out}}$ and $X_{\textrm{in}}\rightarrow X_{\textrm{out}}$. In particular, for the Schwarzschild metric, $K=\frac{1}{4M}$ so that the extra term connecting $t_{\textrm{in}}$ and $t_{\textrm{out}}$ is given by ($-2\pi iM$). Such a result (for the Schwarzschild case) was earlier discussed in \cite{Singleton}.
Now, following the definition (\ref{1.02}), we obtain the relations connecting the null coordinates defined inside and outside the horizon,
\begin{eqnarray}
&&u_{\textrm{in}}=t_{\textrm{in}}-(r_*)_{\textrm{in}} \rightarrow u_{\textrm{out}}-i\frac{\pi}{K}
\nonumber
\\
&&v_{\textrm{in}}=t_{\textrm{in}}+(r_*)_{\textrm{in}} \rightarrow v_{\textrm{out}}
\label{Krus3}
\end{eqnarray}
Under these transformations the inside and outside modes are connected by,
\begin{eqnarray}
&&\Big(\phi^{(R)}\Big)_{\textrm{in}}\rightarrow e^{-\frac{\pi\omega}{\hbar K}}\Big(\phi^{(R)}\Big)_{\textrm{out}}
\nonumber
\\
&&\Big(\phi^{(L)}\Big)_{\textrm{in}}\rightarrow \Big(\phi^{(L)}\Big)_{\textrm{out}}
\label{trans}
\end{eqnarray}
Using the above transformations the density matrix operator for an observer outside the event horizon will be constructed in the next section which will lead to the black body spectrum and thermal flux corresponding to the semiclassical Hawking temperature.

{\textbf{\textit {Black body spectrum and Hawking flux}}}: 
      Now to find the black body spectrum and Hawking flux, we first consider $n$ number of non-interacting virtual pairs that are created inside the black hole. Each of these pairs is represented by the modes defined in the first set of (\ref{1.251}). Then the physical state of the system, observed from outside, is given by,
\begin{eqnarray}
|\Psi> = N \sum_n |n^{(L)}_{\textrm{in}}>\otimes|n^{(R)}_{\textrm{in}}>
 \rightarrow N \sum_n e^{-\frac{\pi n\omega}{\hbar K}}|n^{(L)}_{\textrm{out}}>\otimes|n^{(R)}_{\textrm{out}}>
\label{1.31}
\end{eqnarray}
where  use has been made of the transformations (\ref{trans}). Here $|n^{(L)}_{\textrm{out}}>$ corresponds to $n$ number of left going modes and so on while $N$ is a normalization constant which can be determined by using the normalization condition $<\Psi|\Psi>=1$.  This immediately yields, 
\begin{eqnarray}
N=\frac{1}{\Big(\displaystyle\sum_n e^{-\frac{2\pi n\omega}{\hbar K}}\Big)^{\frac{1}{2}}}
\label{1.32}
\end{eqnarray} 
The above sum will be calculated for both bosons and fermions. For bosons $n=0,1,2,3,....$ whereas for fermions $n=0,1$. With these values of $n$ we obtain the normalization constant (\ref{1.32}) as
\begin{eqnarray}
N_{(\textrm {boson})}=\Big(1-e^{-\frac{2\pi\omega}{\hbar K}}\Big)^{\frac{1}{2}}
\label{1.33}
\\
N_{(\textrm {fermion})}=\Big(1+e^{-\frac{2\pi\omega}{\hbar K}}\Big)^{-\frac{1}{2}}
\label{1.34}
\end{eqnarray}
Therefore the normalized physical states of the system for bosons and fermions are respectively,
\begin{eqnarray}
|\Psi>_{(\textrm{boson})}= \Big(1-e^{-\frac{2\pi\omega}{\hbar K}}\Big)^{\frac{1}{2}} \sum_n e^{-\frac{\pi n\omega}{\hbar K}}|n^{(L)}_{\textrm{out}}>\otimes|n^{(R)}_{\textrm{out}}>
\label{1.35}
\\
|\Psi>_{(\textrm{fermion})}= \Big(1+e^{-\frac{2\pi\omega}{\hbar K}}\Big)^{-\frac{1}{2}} \sum_n e^{-\frac{\pi n\omega}{\hbar K}}|n^{(L)}_{\textrm{out}}>\otimes|n^{(R)}_{\textrm{out}}>
\label{1.36}
\end{eqnarray}
From here on our analysis will be only for bosons since for fermions the analysis is identical. For bosons the density matrix operator of the system is given by,
\begin{eqnarray}
{\hat\rho}_{(\textrm{boson})}&=&|\Psi>_{(\textrm{boson})}<\Psi|_{(\textrm{boson})}
\nonumber
\\
&=&\Big(1-e^{-\frac{2\pi\omega}{\hbar K}}\Big) \sum_{n,m} e^{-\frac{\pi n\omega}{\hbar K}} e^{-\frac{\pi m\omega}{\hbar K}} |n^{(L)}_{\textrm{out}}>\otimes|n^{(R)}_{\textrm{out}}>  <m^{(R)}_{\textrm{out}}|\otimes<m^{(L)}_{\textrm{out}}|
\label{1.37}
\end{eqnarray}
Now tracing out the ingoing (left) modes we obtain the density matrix for the outgoing modes,
\begin{eqnarray}
{\hat{\rho}}^{(R)}_{(\textrm{boson})}= \Big(1-e^{-\frac{2\pi\omega}{\hbar K}}\Big) \sum_{n} e^{-\frac{2\pi n\omega}{\hbar K}}|n^{(R)}_{\textrm{out}}>  <n^{(R)}_{\textrm{out}}|
\label{1.38}
\end{eqnarray}
Therefore the average number of particles detected at asymptotic infinity is given by,
\begin{eqnarray}
<n>_{(\textrm{boson})}={\textrm{trace}}({\hat{n}} {\hat{\rho}}^{(R)}_{(\textrm{boson})})&=& \Big(1-e^{-\frac{2\pi\omega}{\hbar K}}\Big) \sum_{n} n e^{-\frac{2\pi n\omega}{\hbar K}}
\nonumber
\\
&=&\Big(1-e^{-\frac{2\pi\omega}{\hbar K}}\Big) (-\frac{\hbar K}{2\pi})\frac{\partial}{\partial\omega}\Big(\sum_{n}  e^{-\frac{2\pi n\omega}{\hbar K}}\Big)
\nonumber
\\
&=&\Big(1-e^{-\frac{2\pi\omega}{\hbar K}}\Big) (-\frac{\hbar K}{2\pi})\frac{\partial}{\partial\omega}\Big(\frac{1}{1-e^{-\frac{2\pi\omega}{\hbar K}}}\Big)
\nonumber
\\
&=&\frac{1}{e^{\frac{2\pi\omega}{\hbar K}}-1}
\label{1.39}
\end{eqnarray}
where the trace is taken over all $|n^{(R)}_{\textrm{out}}>$ eigenstates. This is the Bose distribution. Similar analysis for fermions leads to the Fermi distribution:
\begin{eqnarray}
<n>_{(\textrm{fermion})}=\frac{1}{e^{\frac{2\pi\omega}{\hbar K}}+1}
\label{1.40} 
\end{eqnarray}
Note that both these distributions correspond to a black body spectrum with a temperature given by the Hawking expression, 
\begin{eqnarray}
T_H=\frac{\hbar K}{2\pi}
\label{1.30}
\end{eqnarray}
Correspondingly, the Hawking flux can be obtained by integrating the above distribution functions over all $\omega$'s. For fermions it is given by,
\begin{eqnarray}
{\textrm{Flux}}=\frac{1}{\pi}\int_0^\infty \frac{\omega ~d\omega}{e^{\frac{2\pi\omega}{\hbar K}}+1} =\frac{\hbar^2 K^2}{48\pi}
\label{1.41}
\end{eqnarray}
Similarly, the Hawking flux for bosons can be calculated, leading to the same answer.

{\textbf{\textit {Connection with usual approaches}}}:  
     For completeness and for revealing the connection with usual approaches \cite{Paddy,Wilczek,Majhi1} to the tunneling formalism we will show below how one can find only the Hawking temperature using the principle of detailed balance.

Since the left moving mode travels towards the center of the black hole, its probability to go inside, as measured by an external observer, is expected to be unity. This is easily seen by computing,   
\begin{eqnarray}
P^{(L)}=|\phi^{(L)}_{\textrm{in}}|^2 \rightarrow|\phi^{(L)}_{\textrm{out}}|^2=1
\label{Krus4}
\end{eqnarray} 
where we have used (\ref{trans}) to recast $\Big(\phi^{(L)}\Big)_{\textrm{in}}$ in terms of $\Big(\phi^{(L)}\Big)_{\textrm{out}}$ since measurements are done by an outside observer. 
This shows that the left moving (ingoing) mode is trapped inside the black hole, as expected.  
     
      On the other hand the right moving mode ($\phi^{(R)}_{\textrm{in}}$) tunnels through the event horizon. So to calculate the tunneling probability as seen by an external observer one has to use the transformation (\ref{trans}) to recast $\Big(\phi^{(R)}\Big)_{\textrm{in}}$ in terms of $\Big(\phi^{(R)}\Big)_{\textrm{out}}$. Then we find, 
\begin{eqnarray}
P^{(R)}=|\phi^{(R)}_{\textrm{in}}|^2 \rightarrow |e^{-\frac{\pi\omega}{\hbar K}}\Big(\phi^{(R)}\Big)_{\textrm{out}}|^2
=e^{-\frac{2\pi\omega}{\hbar K}}
\label{Krus5}
\end{eqnarray}
Finally, using the principle of ``detailed balance'' \cite{Paddy,Majhi2}, $P^{(R)}=e^{-\frac{\omega}{T_H}}P^{(L)}=e^{-\frac{\omega}{T_H}}$ and comparison with (\ref{Krus5}) immediately reproduces the Hawking temperature (\ref{1.30}).

{\textbf{\textit {Conclusions}}}:
     To conclude, we have provided a novel formulation of the tunneling formalism to highlight the role of coordinate systems. A particular feature of this reformulation is that explicit treatment of the singularity in (\ref{1.25}) is not required since we do not carry out the complex path integration. Of course, the singularity at the event horizon is manifested in the transformations (\ref{Krus2}). In this way our formalism, contrary to the traditional approaches \cite{Paddy,Wilczek,Majhi1}, avoids explicit complex path analysis. It is implicit only in the definition (\ref{1.25}). Computations were done in terms of the basic modes. From the density matrix constructed from these modes we were able to directly reproduce the black body spectrum, for either bosons or fermions, from a black hole with a temperature corresponding to the standard Hawking expression. We feel that the lack of such an analysis was a gap in the existing tunneling formulations \cite{Paddy,Wilczek,Majhi1,Majhi2,Majhi3,Majhi4,Singleton, Majhi5} which yield only the temperature rather that the actual black body spectrum. Finally, the connection of our approach with these existing formulations was revealed through the use of the detailed balance principle.


\end{document}